\documentclass[conference]{IEEEtran}
\IEEEoverridecommandlockouts
\usepackage{cite}
\usepackage{amsmath,amssymb,amsfonts}
\usepackage{algorithm}
\usepackage{graphicx}
\usepackage{textcomp}
\usepackage{hyperref}
\usepackage{xcolor}
\usepackage{tikz}
\usepackage{algpseudocode}
\usepackage{algorithm}
\usepackage{algpseudocode}
\usepackage{comment}
\usepackage{enumerate}
\begin{document}

\title{Adaptive, Efficient and Fair Resource Allocation in Cloud Datacenters leveraging Weighted A3C Deep Reinforcement Learning\thanks{Author sequence follows the lexical order of last names.}}

\author{\IEEEauthorblockN{Suchi Kumari}
\IEEEauthorblockA{\textit{Department of Computer Science and Engineering} \\
\textit{Shiv Nadar Institution of Eminence}\\
Delhi-NCR, India \\
\href{mailto:suchi.kumari@snu.edu.in}{suchi.kumari@snu.edu.in}}
\and
\IEEEauthorblockN{Dhruv Mishra}
\IEEEauthorblockA{\textit{Department of Computer Science and Engineering} \\
\textit{Shiv Nadar Institution of Eminence}\\
Delhi-NCR, India \\
\href{mailto:dm409@snu.edu.in}{dm409@snu.edu.in}}}

\maketitle

\begin{abstract}
Cloud data centres demand adaptive, efficient, and fair resource allocation techniques due to heterogeneous workloads with varying priorities. However, most existing approaches struggle to cope with dynamic traffic patterns, often resulting in suboptimal fairness, increased latency, and higher energy consumption. To overcome these limitations, we propose a novel method called Weighted Actor-Critic Deep Reinforcement Learning (WA3C). Unlike static rule-based schedulers, WA3C continuously learns from the environment, making it resilient to changing workload patterns and system dynamics. Furthermore, the algorithm incorporates a multi-objective reward structure that balances trade-offs among latency, throughput, energy consumption, and fairness. This adaptability makes WA3C well-suited for modern multi-tenant cloud infrastructures, where diverse applications often compete for limited resources. WA3C also supports online learning, allowing it to adapt in real time to shifting workload compositions without the need for retraining from scratch. The model’s architecture is designed to be lightweight and scalable, ensuring feasibility even in large-scale deployments. Additionally, WA3C introduces a priority-aware advantage estimator that better captures the urgency of tasks, enhancing scheduling precision. As a result, WA3C achieves more effective convergence, lower latency, and balanced resource allocation among jobs. Extensive experiments using synthetic job traces demonstrate that WA3C consistently outperforms both traditional and reinforcement learning-based baselines, highlighting its potential for real-world deployment in large-scale cloud systems.
\end{abstract}


\section{\textbf{Introduction}}
Cloud data centres are the backbone of modern computing infrastructure, supporting a wide variety of workloads ranging from real-time high-priority tasks, such as online transaction processing and video conferencing, to delay-tolerant batch operations like large-scale data analytics and machine learning model training. This workload heterogeneity, combined with increasing demands for energy efficiency, low latency, and cost-effectiveness, presents major challenges for resource management. Efficient resource allocation in such environments is a key research focus, essential for enhancing system performance, ensuring quality of service (QoS), reducing latency, and minimizing energy usage.

The current state of research has witnessed the emergence of Deep Reinforcement Learning (DRL) techniques as promising alternatives for cloud resource management, offering the ability to learn adaptive policies from interaction with the environment. Models such as Asynchronous Advantage Actor-Critic (A3C) have demonstrated efficiency in general-purpose scheduling scenarios. Nevertheless, existing DRL approaches typically treat all jobs uniformly, failing to account for heterogeneous job priorities. They also lack explicit mechanisms to enforce fairness among competing tasks, which can lead to resource monopolisation by aggressive workloads and the starvation of lower-priority, yet important, tasks. Therefore, in this work, we propose Weighted Asynchronous Advantage Actor-Critic (WA3C), a novel reinforcement learning-based model for cloud resource allocation. The WA3C model enhances the traditional A3C framework by introducing a dynamic reward function that simultaneously incorporates job priority levels and fairness constraints. By adaptively adjusting the reward structure based on workload characteristics, our approach ensures that high-priority tasks receive timely resources while preserving a fair distribution of computational capacity across all jobs. Through extensive experimentation with real-world-inspired datasets, we demonstrate that Weighted A3C significantly outperforms baseline schedulers and standard DRL approaches in terms of efficiency, fairness, and Quality of Service (QoS) compliance.

\begin{table*}[ht]
\centering
\caption{Comparative Research Analysis}
\label{tab:relatedwork}
\begin{tabular}{|p{2.5cm}|p{5cm}|p{2.5cm}|p{2.5cm}|p{3cm}|}
\hline
\textbf{Reference Article} & \textbf{Approach} & \textbf{Time Complexity} & \textbf{Memory Usage} & \textbf{Scalability} \\ \hline
Xu \textit{et al.} ~\cite{xu2013dynamic} & Online adaptive scheduling for elastic services using heuristic-based optimization & Low to Moderate & Low & Limited scalability to large clusters \\ \hline
Mao \textit{et al.} ~\cite{mao2016resource} & DRL for resource management (DeepRM) using policy gradient methods & High & High (deep RL overhead) & Moderate (single-agent learning) \\ \hline
Lu \textit{et al.} ~\cite{lu2024a2c} & Advantage Actor-Critic (A2C) Learning with convergence to PPO behavior & Moderate & Moderate (lighter than A3C) & Moderate (single-agent learning) \\ \hline
Mnih \textit{et al.} ~\cite{mnih2016asynchronous} & Asynchronous Advantage Actor-Critic (A3C) learning & Moderate to High & Moderate & High (asynchronous, multi-agent scalability) \\ \hline
Chen \textit{et al.} ~\cite{chen2021adaptive} & A3C-based adaptive resource allocation using MDP formulation & Moderate & Moderate (deep model overhead) & High (parallel learning and MDP adaptability) \\ \hline
\textbf{Proposed \textit{WA3C} Model} & Priority-aware Weighted A3C with fairness-enhanced dynamic reward adjustment & Moderate & Moderate (optimized overhead) & High (supports dynamic workloads) \\ \hline
\end{tabular}
\end{table*}

The primary objectives of this research are as follows:

\begin{itemize}
    \item To develop a Weighted Asynchronous Advantage Actor-Critic (WA3C) scheduling algorithm that incorporates job priorities and fairness constraints into the reinforcement learning framework for efficient cloud resource management.
    
    \item To design a dynamic and composite reward function that balances Quality of Service (QoS) metrics, job priority levels, energy efficiency, and system fairness.
    
    \item To prevent resource monopolisation by aggressive workloads and ensure fair resource allocation, thereby avoiding starvation of lower-priority tasks.
    
    \item To evaluate the performance of the proposed WA3C approach against baseline scheduling algorithms and standard DRL methods using real-world-inspired datasets.
\end{itemize}

The remainder of the manuscript is organised as follows. Section \ref{Sec2} reviews the related research work considering classical, machine learning, and deep learning based models for resource allocation in a cloud computing domain. Section \ref{Sec3} describes the architecture and details of the proposed Weighted Asynchronous Advantage Actor Critic (\textit{WA3C}) model. The environmental setup, the description of the data set, the results, and the in-depth analysis are presented in Section \ref{Sec4}. Section \ref{Sec5} summarizes the efficacy of the proposed \textit{WA3C} model, practical implications, and its limitations. Finally, Section \ref{Sec6} concludes the paper and outlines the future direction.

\section{\textbf{Related Work}} \label{Sec2}
 Traditional approaches, such as heuristic-based scheduling policies \cite{xu2013dynamic, sharma2011balancing} and static resource provisioning models \cite{ghosh2015dynamic}, have provided lightweight solutions but often fail to adapt to dynamic and heterogeneous workloads \cite{prasad2025enhancing}. Wei \textit{et al.} \cite{wei2010game} addresses QoS-constrained resource allocation using a game theory approach. A two-step approach is provided: independent Binary Integer Programming optimisation followed by an evolutionary mechanism ensuring fairness and convergence to a Nash equilibrium. Ghanbari \textit{et al.} \cite{ghanbari2019resource} present a literature review of resource allocation methods in IoT, categorising approaches based on parameters like cost, context, efficiency, quality of service, and service level agreement. Belgacem \textit{et al.} \cite{belgacem2020dynamic} introduce a dynamic resource allocation model using the Multi-Objectives Symbiotic Organism Search (MOSOS) to optimise makespan, QoS, and cost in cloud environments. Murad \textit{et al.} \cite{murad2024sg} proposed a priority-based fair scheduling algorithm to assign optimal time to finish the jobs using the resources in the cloud. They applied backfilling strategies to minimise the gap in scheduling the job on the cloud.

Early machine learning techniques, including supervised learning models for workload prediction \cite{shi2011learning,bi2021integrated} and queuing theory-based models \cite{urgaonkar2005resource}, introduced adaptive decision-making but lacked real-time learning capabilities. Lattuada \textit{et al.} \cite{lattuada2020optimal} propose intelligent resource allocation policies for Spark applications in cloud environments, ensuring QoS under hard and soft deadlines. Machine learning techniques helped in achieving lower prediction errors and faster resource rebalancing compared to existing methods. Shahidinejad \textit{et al.} \cite{shahidinejad2021resource} provided a hybrid resource provisioning approach by clustering workloads and applying decision trees for scaling decisions to incorporate load balancing in the cloud environment. Khodaverdian \textit{et al.} \cite{khodaverdian2024energy} provided a hybrid model considering a Convolutional Neural Network (CNN) and a Gated Recurrent Unit (GRU). The hybrid approach is capable of classifying VMs as Delay-Sensitive or Delay-Insensitive to optimise live migration in cloud datacenters, using workload prediction. Reinforcement learning (RL) methods, such as Q-learning-based resource allocators \cite{mao2016resource}, demonstrated more promising results by enabling agents to learn directly from interactions. Djigal \textit{et al.} \cite{djigal2022machine} surveyed multiple Machine Learning and Deep Learning based approaches for resource allocation (RA) in Multi-Access Edge Computing. Zhou \textit{et al.} \cite{zhou2022aquatope} introduced a quality of service and uncertainty-aware scheduler (Aquatope) for multi-stage serverless workflows that leverages Bayesian models to prewarn containers and allocate resources efficiently. However, these methods faced challenges such as slow convergence, inability to generalise to unseen workload patterns, and scalability issues in high-dimensional action spaces. 

More recently, deep reinforcement learning (DRL) methods have gained attention for cloud resource management. Sutton \textit{et al.} \cite{sutton1998reinforcement} proposed an Advantage Actor-Critic (A2C) model, where the actor selects actions based on feedback from the critic. Actor-Critic methods, particularly A2C frameworks, have shown notable improvements in both training stability and policy optimisation \cite{mnih2016asynchronous}. However, their approach uses single-threaded training, with only one DRL agent interacting with the environment. This leads to underutilization of computational resources and limits overall performance. Mnih \textit{et al.} \cite{mnih2016asynchronous} addressed these problems and proposed a new model named Asynchronous Advantage Actor-Critic (A3C), which used multiple DRL agents for interacting with the environment with reduced latency and enhanced resource utilisation. Chen \textit{et al.} ~\cite{chen2021adaptive} proposed an A3C-based adaptive resource allocation strategy that significantly outperformed traditional and earlier RL techniques, demonstrating improved job scheduling efficiency and energy savings. Chen \textit{et al.} \cite{chen2022resource} used a Deep Reinforcement Learning-based resource allocation method that considers both current and future workloads using workload-time windows. This scheme achieves high prediction accuracy and outperforms traditional methods in dynamic cloud environments. Zhou \textit{et al.} \cite{zhou2024deep} present a comprehensive review of Deep Reinforcement Learning (DRL) methods for resource scheduling in cloud computing. However, existing DRL models often treat all jobs equally, neglecting job-specific priorities and fairness considerations, which are critical in real-world multi-tenant cloud environments. To address these gaps, our work extends the A3C framework by introducing a priority-aware and fairness-constrained reward mechanism, termed Weighted A3C (WA3C), thus enhancing both the adaptability and equity of resource allocation in dynamic cloud settings.

\section{\textbf{Proposed Approach}} \label{Sec3}
In this work, we propose a novel extension of the Asynchronous Advantage Actor-Critic (A3C) framework, which we term Weighted A3C (WA3C). The original A3C algorithm is a reinforcement learning (RL) paradigm that leverages multiple parallel agents (workers) to asynchronously update a global neural network, consisting of an actor (which learns a policy $\pi(a|s)$) and a critic (which estimates a value function $V(s)$). The actor selects actions based on the policy, while the critic provides feedback on the action’s quality using the advantage estimate. This design achieves both scalability and stability during training.

Our proposed WA3C model advances the baseline A3C architecture by integrating critical real-world constraints pertinent to cloud computing environments, specifically job prioritisation, energy, fairness enforcement, and quality of service metrics. These enhancements ensure that the scheduling policy not only prioritises urgent jobs but also prevents starvation of lower-priority tasks and enforces equitable resource distribution across users and job types. Improvements are realised through three core innovations discussed as follows.

\begin{enumerate}
    \item \textbf{Composite Reward Function:} Unlike conventional A3C that may use simple scalar rewards (e.g., based on latency), the proposed \textit{WA3C} introduces a multi-objective reward function that assigns dynamic weights to various system-level objectives. These include job priority, fairness (measured via metrics such as Jain's index), dismissal penalties (for dropped or late jobs), quality of service (QoS), and energy efficiency. This guides the learning agent to optimise both system performance and equitable resource distribution.
    \item \textbf{Priority and Fairness-Aware Policy:} The action selection policy is enhanced using a priority-weighted softmax mechanism:
    \[\pi(a \mid s) \propto \exp(Q(a, s) + \beta \cdot \mathcal{P}(a))
    \]
    where \( Q(a, s) \) is the estimated action-value function, \( \mathcal{P}(a) \) represents the priority or urgency of the job, and \( \beta \) is a tunable hyperparameter that adjusts the influence of the priority term. This ensures that high-priority jobs are scheduled promptly while also preserving fairness among competing tasks and mitigating unnecessary dismissals.
    \item \textbf{TD Error and Adaptive Actor-Critic Learning:} The model uses Temporal Difference (TD) error to update both the actor and critic networks:
    \[\delta_t = R_t + \gamma V(s_{t+1}) - V(s_t)\]
    This allows for continuous, online learning in a dynamic cloud environment. The TD error captures the discrepancy between expected and actual outcomes, guiding both policy and value updates. As a result, the model makes better long-term scheduling decisions that balance priority, fairness, and efficiency while minimising the rate of dismissed jobs.
\end{enumerate}

\begin{figure}[htp]
    \centering
    \includegraphics[width=0.5 \textwidth, height = 2 in]{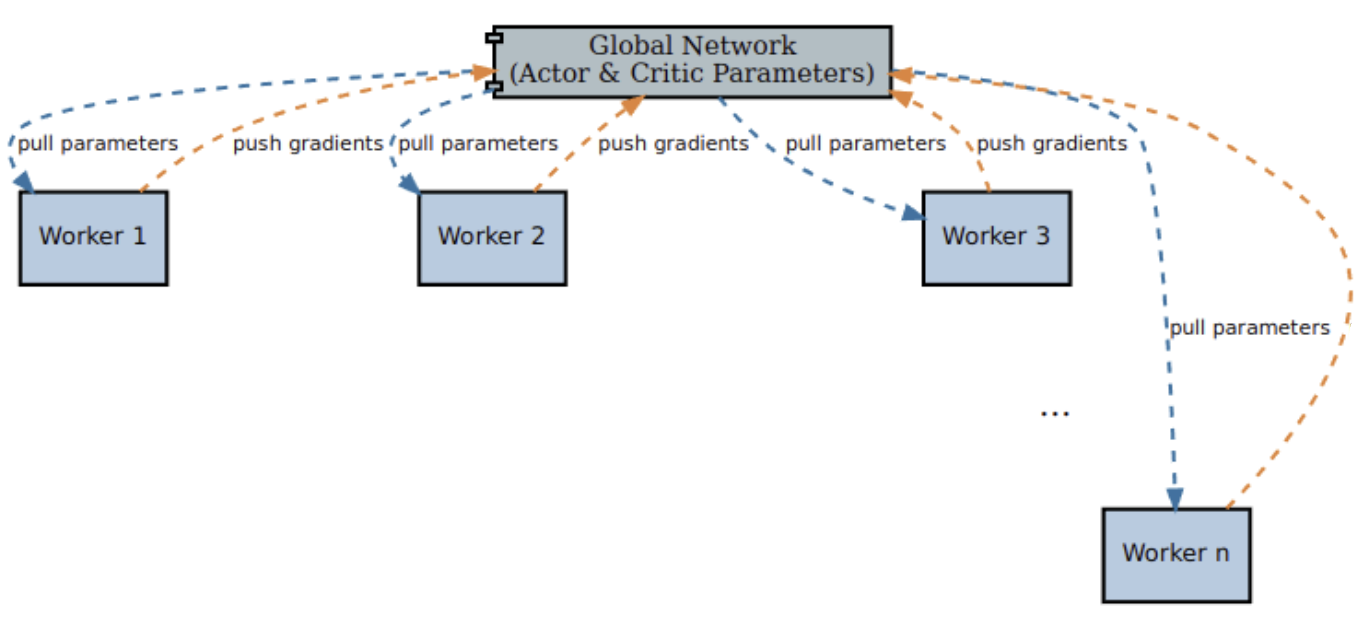}
    \caption{n-Workers collaboration in WA3C model}
    \label{fig:n_workers}
\end{figure}

In the \textit{WA3C} framework, the training process is distributed across \( n\) parallel worker agents, each interacting independently with its own instance of the cloud environment. These worker agents asynchronously collect experiences (i.e., states, actions, rewards, next states) and compute gradients with respect to the local copy of the policy and value networks. Periodically, the gradients are pushed to a shared \textit{global network}, which maintains the central model parameters. The global network then updates its parameters using these gradients and sends the updated weights back to the worker agents, enabling continual synchronisation. This setup enables high scalability and faster convergence, as depicted in \textbf{Figure \ref{fig:n_workers}}, which illustrates how \( n \) workers operate concurrently and communicate with the central global network. Here, \( n \) corresponds to the number of available Virtual Machines (VMs), while the total number of jobs in the system is denoted by \( {N} \).

\begin{figure*}[htp]
    \centering
    \includegraphics[width=0.9 \linewidth, height=4.5 in]{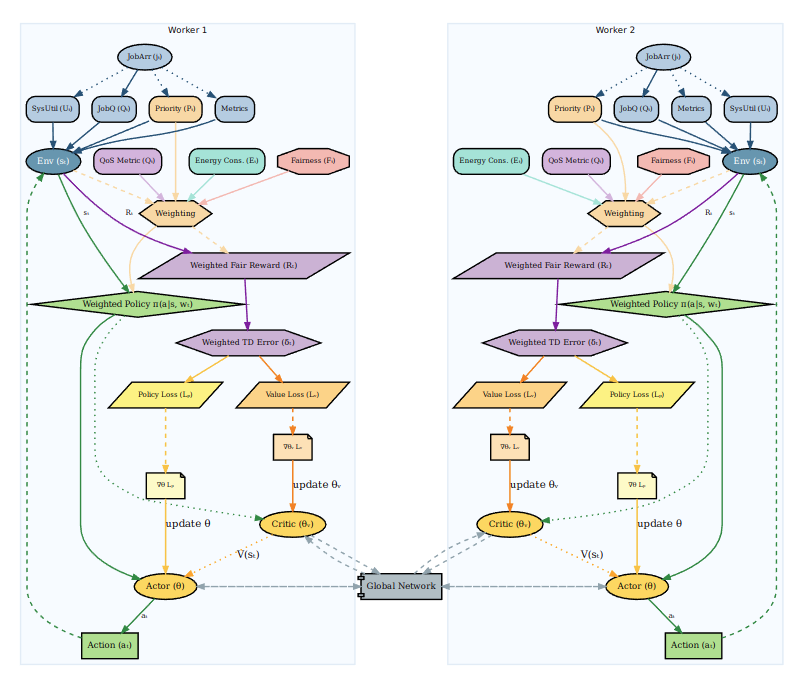}
    \caption{Detailed Two Worker collaboration in WA3C model}
    \label{fig:2_workers}
\end{figure*}

\textbf{Figure \ref{fig:2_workers}} provides a more detailed breakdown of the process with two workers. Each worker observes its environment state, selects actions based on its local policy, computes the \textit{TD} error, and performs backpropagation to compute gradients. These gradients are then asynchronously sent to the global network, which updates the global parameters using an optimisation algorithm such as RMSProp/Adam \cite{zou2019sufficient}. The workers pull back the updated global parameters, allowing them to benefit from the collective learning of all agents. This asynchronous mechanism not only ensures robustness to delays and stragglers but also leads to better generalisation and adaptability in dynamic cloud environments.

\subsection{\textbf{System Model}}

The job scheduling and resource allocation problem is formulated as a Markov Decision Process (MDP) defined by the tuple $⟨S,A,R,P,\gamma⟩$, where the agent learns to optimise long-term performance using the proposed \textit{WA3C} framework, enriched with priority-aware and fairness-constrained rewards. Here, the system contains $j \in N$ jobs, where $N$ denotes the set of all jobs.

\begin{itemize}
    \item \textbf{State Space} (\(S\)): Each state \(s_t \in S\) represents a comprehensive snapshot of the system at time \(t\). The state is composed of a rich set of features that capture both the current status of the system and the characteristics of the job. It can be written as follows.
    \[
    s_t = \{ Q_t, U^{Resource}_t, \mathcal{P}_t, U^{CPU}_t, \text{MAPI}_t, \text{CPI}_t \}
    \]
    where \(Q_t\) represents the current job queue containing metadata for each job \(j\), such as estimated runtime, resource demand, arrival time, and deadline. \(U^{Resource}_t\) shows system resource utilisation statistics, including CPU load, memory usage, and I/O throughput. \(\mathcal{P}_t\) is the priority vector encoding the importance level of each job, typically derived from policy or SLA requirements. \(U^{CPU}_t\) shows current CPU usage percentage, normalised to [0, 1], indicating the workload pressure on the processor. \(\text{MAPI}_t\) is the Memory Accesses Per Instruction, reflecting the memory-intensiveness of the workload and \(\text{CPI}_t\) represents Cycles Per Instruction, capturing the computational efficiency of the current job mix.

    This detailed state representation allows the actor-critic model to learn context-aware scheduling decisions that are sensitive to workload diversity, job importance, and system bottlenecks.

\item \textbf{Action Space} (\(A\)):  
At each timestep, the agent selects an action \(a_t \in A\), where each action corresponds to scheduling a specific job \(j\) from the job queue \(Q_t\). Mathematically, it can be represented as follows.
\[
A = \{ a_t \mid a_t = \text{select job } j \in Q_t \}
\]
To incorporate priority into the policy, the actor uses a softmax distribution over the estimated \textit{Q-}values adjusted by a weighted priority score \(\mathcal{P}_j\).
\[
\pi(a|s) \sim \text{softmax}(Q(s, j) + \beta \cdot \mathcal{P}_j)
\]
This design ensures that higher-priority jobs are more likely to be selected, while still allowing exploration.

\item \textbf{Transition Probability} (\(P\)):  
The environment is stochastic, and the transition function
\[
P(s_{t+1} | s_t, a_t)
\]
models how the system evolves after a job is scheduled. Changes in resource usage, queue dynamics, and job completions are captured here.

\item \textbf{Discount Factor} (\(\gamma\)):  
A fixed discount factor \( \gamma = 0.95 \) is used to balance immediate and future rewards, encouraging long-term optimisation without sacrificing responsiveness.

\end{itemize}

In this formulation, the actor network is responsible for learning the job selection policy $\pi(a|s)$, while the critic network estimates the value function to guide policy improvement. The integration of job priorities and fairness ensures that the system is both efficient and equitable, aligning with QoS guarantees and energy objectives. 

\subsection{\textbf{Weighted Reward Function}}

At each time step \( t \), the agent receives a total reward \( R_t \) defined in Eq. \eqref{eq:reward}.

\begin{equation}
R_t = w_1 R_t^{QoS} + w_2 R_t^{\mathcal{E}} + w_3 R_t^{\mathcal{P}} + w_4 R_t^{\mathcal{F}} + w_5 R_t^{\mathcal{D}}. \label{eq:reward}
\end{equation}

The hyperparameters $w_1 = 0.25, w_2 = 0.2, w_3 = 0.25, w_4 = 0.15, w_5 = 0.15$ are used to balance the contribution of each reward component. These weights are not uniformly distributed, as their values depend on the specific application context. For example, in network-oriented applications where fair resource allocation is critical, a higher value is typically assigned to $w_4$. In contrast, real-time systems prioritise timely responses, giving more importance to $w_3$. In scenarios where data loss is unacceptable, $w_5$ is assigned a larger weight. The weights are normalised such that $\sum_i w_i = 1$. Now, each reward component is explained in detail in the following section.

\subsubsection{\textbf{Quality of Service Reward (\( R_t^{QoS} \))}}

The QoS reward incentivises minimising job latency and is defined in Eq. \eqref{eq:QoSreward}.

\begin{equation}
R_t^{QoS} = 1 - \frac{L_j}{L_{\max}} \label{eq:QoSreward}
\end{equation}

where \( L_j \): Actual latency experienced by job \( j \) and \( L_{\max} \): Maximum tolerable latency in the system. 

The job latency \( L_j \) is formulated considering all types of delays incurred before assigning the job to the system. It is formulated in Eq. \eqref{eq:latency}.

\begin{equation}
L_j = T^{wait}_j + T^{queue}_j + T^{exec}_j \label{eq:latency}
\end{equation}

where \( T^{wait}_j \) is the time spent waiting for resources to be available, \( T^{queue}_j \)is the time spent in the scheduling queue, and \( T^{exec}_j \) is the actual execution time of the job on the assigned resource. Here we assume
$$
T^{exec}_j = \frac{C_j}{f \cdot (1 - \theta_j)}
$$
where \( C_j \) is the total computation units required by job \( j \), \( f \) is the CPU frequency (in GHz), and \( \theta_j \) is the interference factor due to other co-located jobs (set to 0.2 based on empirical tuning).

\subsubsection{\textbf{Energy Efficiency Reward (\( R_t^{\mathcal{E}} \))}}

To promote sustainability, the energy reward penalises high energy consumption and is expressed in Eq. \eqref{eq:energyreward}.

\begin{equation}
R_t^{\mathcal{E}} = -\alpha \times \mathcal{E}_j \label{eq:energyreward}
\end{equation}

where \( \mathcal{E}_j \) is the normalized energy consumption for job \( j \) and \( \alpha \) is energy penalty scaling constant.

To accurately account for the energy efficiency of the resource allocation policy, we introduce an enhanced model for estimating energy consumption based on system-level utilisation metrics. Unlike traditional static models, this formulation dynamically incorporates both computational intensity and memory access patterns of executing jobs. The energy consumption $\mathcal{E}_j$ for a job $j$ is calculated using the following formula in Eq. \eqref{eq:energyjob}.

\begin{equation}
\mathcal{E}_j = \left( \mathcal{W}_{\text{base}} + U^{CPU} \times \iota
 \times (\mathcal{W}_{\text{max}} - \mathcal{W}_{\text{base}}) \right) \times \tau \label{eq:energyjob}
\end{equation}

where \( \mathcal{W}_{\text{base}} \) is the baseline power consumption of the system (e.g., $100$ Watts), \( \mathcal{W}_{\text{max}} \): Maximum power consumption under full load (e.g., \textit{200} Watts),\( U^{CPU} \) is the fraction of CPU usage ( normalized to the range \([0, 1]\)),\(\tau  \) is the execution time of the job in seconds, and \( \iota
 \) is the dynamic performance coefficient capturing workload characteristics.

The performance coefficient \( \iota \) is dependent on the CPU and memory access by each instruction. It is evaluated in Eq. \eqref{eq:iota}.

\begin{equation}
\iota = \text{CPI} \times z + \text{MAPI} \times (1 - z) \label{eq:iota}
\end{equation}

where \( \text{CPI} \) is the Cycles Per Instruction, indicating computational complexity. \( \text{MAPI} \) is the Memory Accesses Per Instruction, indicating memory intensity, and \( z \) is the tunable hyperparameter (set to $0.3$) controlling the influence of memory access vs. computational complexity.

\subsubsection{\textbf{Priority Satisfaction Reward (\( R_t^{\mathcal{P}} \))}}

The priority reward ensures that jobs with higher importance are served more responsively and is formulated in Eq. \eqref{eq:priorityreward}.

\begin{equation}
R_t^{\mathcal{P}} = \mathcal{P}_j \times (1 - L_j) \label{eq:priorityreward}
\end{equation}

where \( \mathcal{P}_j \) is the normalized priority score of job \( j \) in the range \([0, 1]\), and \( L_j \) is the latency experienced by job \( j \) (explained in Eq. \eqref{eq:latency}).
\medskip

\subsubsection{\textbf{Fairness Reward (\( R_t^{\mathcal{F}} \))}}

Fairness is captured by penalising the variance in resource utilisation across all jobs. A lower variance indicates a more equitable allocation of resources, suggesting that the jobs are fairly distributed within the cloud environment. The corresponding formulation is presented in Eq. \eqref{eq:rewardfairness}.

\begin{equation}
R_t^{\mathcal{F}} = -\lambda \times \sigma^2(U^{Resource}) \label{eq:rewardfairness}
\end{equation}

where \( \sigma^2(U^{Resource}) \) is the variance in resource utilization vector \( U^{Resource} \) across active jobs and \( \lambda \) is the regularization factor penalizing unfairness in resource allocation.

\subsubsection{\textbf{Dismissal Penalty Reward (\( R_t^\mathcal{D} \))}}

When the system is overloaded,low priority jobs may be dismissed due to insufficient available resources. However, once a job is dismissed, it cannot be recovered. To account for this, we impose a penalty on excessive job rejections. The corresponding formulation is given in Eq. \eqref{eq:rewarddismissal}.

\begin{equation}
R_t^\mathcal{D} = -\mu \times D_t \label{eq:rewarddismissal}
\end{equation}

where \( D_t \): Number of jobs dismissed at time \( t \) and \( \mu = 0.5 \): Penalty scaling constant for job dismissals.

\subsection{\textbf{Priority-Weighted Softmax Action Selection}}

To ensure that urgent and high-priority jobs are scheduled promptly without compromising overall system fairness, our model modifies the conventional action selection mechanism. Instead of relying solely on estimated rewards, the action selection incorporates job priority directly into the decision-making process. The agent selects the next job $j$ to schedule at time $t$ using a priority-weighted softmax function $\mathcal{P}(a_t = j)$ over the estimated action-value function $Q(s_t, j)$. The formulation is provided in Eq. \eqref{eq:softmax}.

\begin{equation}
\mathcal{P}(a_t = j) = \frac{e^{(Q(s_t, j) + \beta \times \mathcal{P}_j)}}{\sum_{k} e^{(Q(s_t, k) + \beta \times \mathcal{P}_k)}} \label{eq:softmax}
\end{equation}

where \( Q(s_t, j) \) is the action-value function estimating the utility of selecting job \( j \) in state \( s_t \), \( \mathcal{P}_j \): Priority score of job \( j \), and \( \beta = 2.0 \) is the scaling factor that amplifies the influence of job priorities during action selection.

\subsection{\textbf{Algorithm Description}}
This section presents the job scheduling strategies of the proposed WA3C model (Algorithm \ref{algo1}), along with the asynchronous policy parameter updates for each DRL agent (Algorithm \ref{algo2}).

\begin{algorithm}[ht]
\caption{WA3C-Based Job Scheduling in Cloud Environments}
\begin{algorithmic}[1]
\State \textbf{Initialize:} Global actor network $A_\theta$ and critic network $C_\phi$ with weights $\theta$ and $\phi$
\State \textbf{Initialize:} Learning rates $\alpha_\theta$, $\alpha_\phi$, reward decay $\gamma$, TD discount $\beta$, step counter \texttt{temp} $\gets 0$, sync interval $u$
\For{each training epoch $n = 0, 1, 2, \ldots, N$}
    \State Receive initial state $s_0$ from environment: $s_0 \gets \texttt{env.observe()}$
    \For{$t = 0, 1, 2, \ldots, T$}
        \State Form state vector $s_t = [s_t^J, s_t^S]$ (job queue, system load, etc.)
        \State Select scheduling action $a_t$ using modified softmax: 
        \State \hspace{1em} $a_t \gets \texttt{actor.select}(s_t; \alpha)$
        \State Execute $a_t$, observe reward $R_t$ and next state $s_{t+1}$:
        \State \hspace{1em} $s_{t+1}, R_t \gets \texttt{env.step}(a_t)$
        \State Reward is computed as: 
        \State \hspace{1em} $R_t = w_1 R_t^{QoS} + w_2 R_t^{\mathcal{E}} + w_3 R_t^{\mathcal{P}} + w_4 R_t^{\mathcal{F}} + w_5 R_t^{\mathcal{D}}$
        \State Compute discounted return:
        \State \hspace{1em} $R^{\text{d}} = R_t + \gamma R_{t+1} + \ldots + \gamma^k R_{t+k}$
        \State Estimate advantage:
        \State \hspace{1em} $A(s_t, a_t) = R^{\text{d}} + \beta V_\phi(s_{t+1}) - V_\phi(s_t)$
        \State Compute TD error:
        \State \hspace{1em} $\delta_t = R_t + \beta V_\phi(s_{t+1}) - V_\phi(s_t)$
        \State Update critic parameters:
        \State \hspace{1em} $\phi \gets \phi + \alpha_\phi \delta_t \nabla_\phi V_\phi(s_t)$
        \State Compute policy gradient:
        \State \hspace{1em} $\nabla_\theta J(\theta) = \mathbb{E}[\nabla_\theta \log \pi_\theta(s_t, a_t) \cdot A(s_t, a_t)]$
        \State Update actor parameters:
        \State \hspace{1em} $\theta \gets \theta + \alpha_\theta \nabla_\theta J(\theta)$
        \State Update state: $s_t \gets s_{t+1}$
        \State Increment counter: $\texttt{temp} \gets \texttt{temp} + 1$
        \If{$\texttt{temp} \bmod u = 0$}
            \State \textit{Push gradients} to global network and \textit{pull latest parameters} (see Algorithm~2)
        \EndIf
    \EndFor
\EndFor
\end{algorithmic} \label{algo1}
\end{algorithm}

\subsubsection{Algorithm 1: WA3C-Based Job Scheduling in Cloud Environments}

This algorithm presents the training workflow of the proposed WA3C  model applied to job scheduling in a cloud computing environment (in Algorithm \ref{algo1}). The system begins by initialising the actor and critic networks with parameters $\theta$ and $\phi$, respectively. The learning process proceeds over a series of training epochs. At each time step $t$, the current state $s_t$ is constructed using various system metrics (e.g., job queue state $s_t^J$, system load $s_t^S$). The actor selects a scheduling action $a_t$ based on a modified softmax policy.

Upon executing the action, the agent receives a reward $R_t$ that reflects multiple objectives, such as Quality-of-Service (QoS), energy efficiency, job priority, fairness, and penalties for constraint violations. The total return is computed through discounted cumulative rewards. The advantage function is estimated to gauge the value of the selected action relative to the baseline value function $V_\phi$. The temporal difference (TD) error $\delta_t$ is used to update the critic using gradient ascent, while the policy gradient, weighted by the advantage, is used to update the actor. Periodically (every $u$ steps), the local parameters synchronise with the global network using the asynchronous update routine described in Algorithm \ref{algo2}.

\begin{algorithm}[ht]
\caption{Asynchronous Update of Policy Parameters in Each DRL Agent}
\begin{algorithmic}[1]
\State \textbf{Initialize:} Global actor parameters $\theta$, local actor parameters $\theta'$, global critic parameters $\phi$, local critic parameters $\phi'$
\State \textbf{Initialize:} Empty gradients $d\theta \gets 0$, $d\phi \gets 0$
\For{$i = \texttt{temp} - u$ to $\texttt{temp}$}
    \State Accumulate actor gradients:
    \State \hspace{1em} $d\theta \gets d\theta + \nabla_{\theta'} \log \pi_{\theta'}(s_i, a_i) \cdot (R_i - V_\phi(s_i))$
    \State Accumulate critic gradients:
    \State \hspace{1em} $d\phi \gets d\phi + \nabla_{\phi'} (R_i - V_{\phi}(s_i))^2$
\EndFor
\State Update global parameters via RMSProp:
\State \hspace{1em} $\theta \gets \theta + \alpha_\theta \cdot d\theta$
\State \hspace{1em} $\phi \gets \phi + \alpha_\phi \cdot d\phi$
\State Synchronise local parameters:
\State \hspace{1em} $\theta' \gets \theta$, $\phi' \gets \phi$
\State Reset gradients: $d\theta \gets 0$, $d\phi \gets 0$
\end{algorithmic} \label{algo2}
\end{algorithm}

\subsubsection{Algorithm 2: Asynchronous Update of Policy Parameters in Each DRL Agent}
This algorithm performs the synchronisation of local and global networks in the proposed \textit{WA3C} setup. Each agent collects experiences over $u$ steps and accumulates gradients for both actor and critic networks. The actor gradient is computed as the product of the log-probability of the action and the advantage $(R_i - V_\phi(s_i))$. The critic gradient corresponds to minimising the squared temporal difference error. After accumulating gradients over $u$ steps, the global parameters $\theta$ and $\phi$ are updated using RMSProp is an adaptive learning rate optimisation algorithm commonly used to train neural networks, especially in reinforcement learning and deep learning tasks.

These updated parameters are then pulled by the local networks to stay in sync. This method enables scalable, asynchronous reinforcement learning where multiple agents explore different parts of the environment in parallel, promoting better generalisation and faster convergence.

\begin{algorithm}[ht]
\caption{Victim Selection for Resource Management}
\label{alg:victim}
\begin{algorithmic}[1]
\State \textbf{Input:} Job queue \( Q_t \), priority set \( \mathcal{P}_t \), threshold \( T_{\text{max}} \)
\If{Current resource usage \( > T_{\text{max}} \)}
    \State Sort jobs in \( Q_t \) by ascending priority \( \mathcal{P}_j \)
    \While{Resource usage \( > T_{\text{max}} \)}
        \State Dismiss job with lowest priority from \( Q_t \)
        \State Increment dismissal counter: \( \mathcal{D}_t \gets \mathcal{D}_t + 1 \)
    \EndWhile
\EndIf
\end{algorithmic} \label{algo3}
\end{algorithm}

In Algorithm \ref{algo2}, the victims are selected using the steps in Algorithm \ref{algo3}. The utility algorithm is invoked when the system's resource usage exceeds a predefined threshold $T_{max}$. It takes as input the current job queue $Q_t$ and the associated priority values $\mathcal{P}_t$. The jobs are sorted in ascending order of priority, and the lowest-priority jobs are dismissed one by one until the resource usage falls below $T_{max}$. Each dismissal increments a counter $\mathcal{D}_t$, which can be used to track the impact of over-utilisation management. This mechanism ensures graceful degradation by preserving high-priority jobs during overload conditions.

\section{\textbf{Results and Analysis}} \label{Sec4}
This section describes the experimental setup, including the environment, dataset, and performance metrics used to evaluate the proposed model. It also provides the visual analysis and comparative results with the existing approaches.

\subsection{\textbf{Experimental Setup and Dataset}}
The experiments were conducted on a 2-socket Intel Xeon CPU \textit{E5-2690 v4} machine, equipped with \textit{32} cores per socket, running Ubuntu \textit{20.04} LTS. Synthetic workloads consisting of \textit{1000} jobs with varying priority levels were generated to simulate realistic cloud datacenter conditions. Additionally, real-world job traces were incorporated from the publicly available \textbf{\textit{Google Cluster Workload Dataset}}\cite{reiss2011googlecluster}, which contains over \textit{29} days of job submissions from a production-scale cluster consisting of more than \textit{11,000} machines. The dataset includes detailed logs of task events, job resource usage, priority levels, scheduling class, and machine-level statistics such as CPU, memory, and disk usage. This rich dataset enabled robust validation of the model under authentic arrival patterns, heterogeneous resource demands, fluctuating workloads, and varying job durations observed in large-scale cloud environments. Both the baseline Standard A3C and the proposed \textit{WA3C} models were trained for \textit{500 }episodes.

\subsection{\textbf{Performance Metrics}}
To evaluate the \textit{WA3C} framework, we adopt a comprehensive set of metrics and controlled comparisons to capture both instantaneous and long-term performance dynamics across different scheduling strategies and learning methods. Performance metrics, including raw rewards, moving average rewards, and aggregate statistics such as average job completion time, job dismissal rate, energy efficiency, and fairness index, were evaluated to assess the effectiveness of the model in both synthetic and real-world conditions. These metrics jointly evaluate both the effectiveness (performance peaks) and the stability (variance and consistency) of each approach.

\subsection{\textbf{Analysis of Sensitivity to Discount Factor and Learning Rate:}}
We analyse the sensitivity of the WA3C framework to its two most critical hyperparameters: the discount factor \((\beta)\) and the learning rate \((\gamma)\). These parameters significantly impact the agent’s emphasis on long-term versus short-term rewards and the stability of the learning process.
\medskip
\subsubsection{\textbf{Effect of Discount Factor \((\beta)\) on \textit{WA3C} framework}}

\begin{figure}[htp]
    \centering
    \includegraphics[width=0.5 \textwidth, height = 2 in]{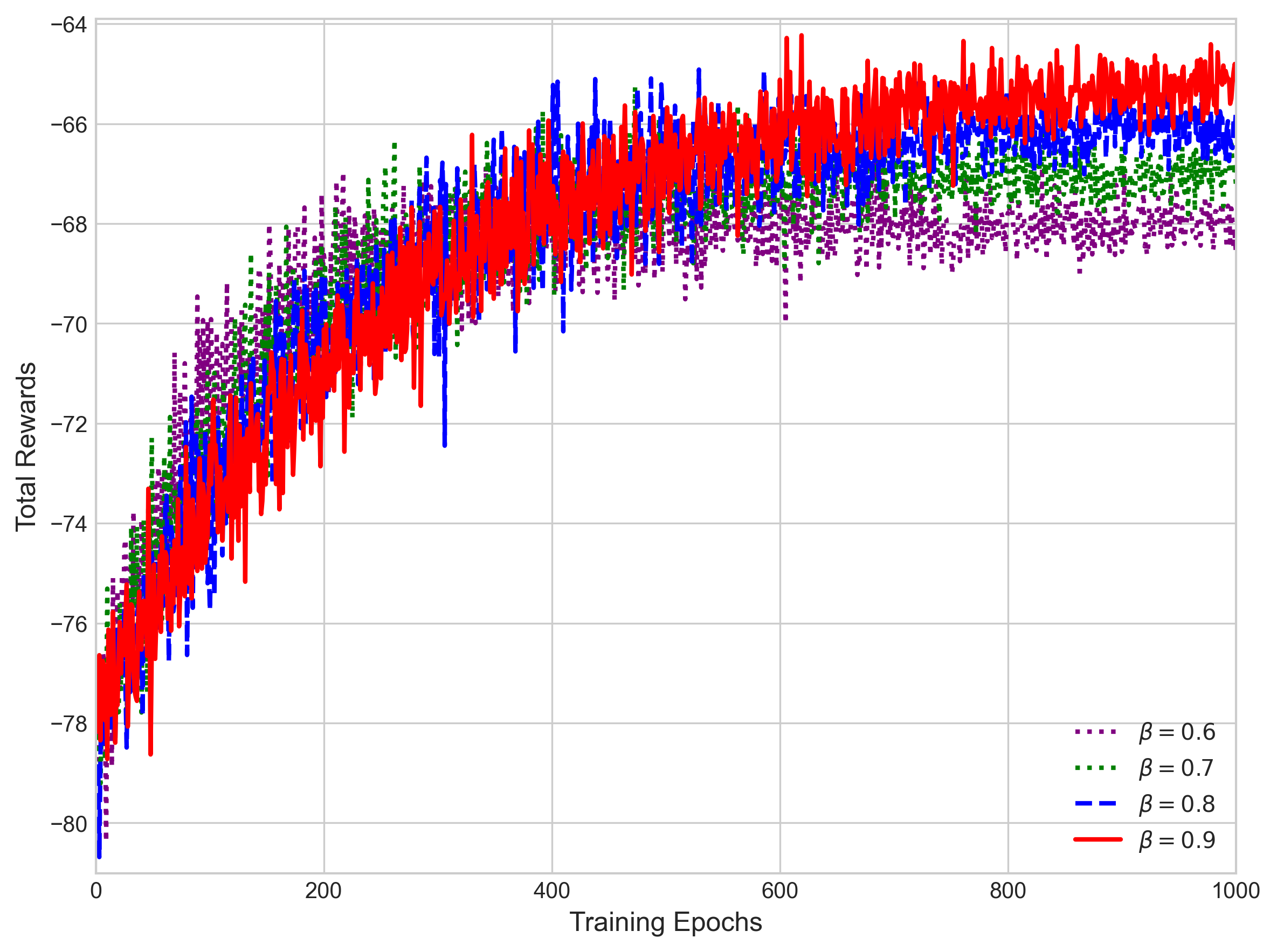}
    \caption{Sensitivity of the proposed \textit{WA3C} model to the varying Discount Rates}
    \label{fig:discount_factors}
\end{figure}

As illustrated in \textbf{Figure~\ref{fig:discount_factors}}, we vary the discount factor across $\beta = \{0.6, 0.7, 0.8, 0.9\}$  to examine its impact on total rewards over training epochs. A higher discount factor, such as $\beta = 0.9$ (solid red line), consistently achieves the highest total rewards, converging around $-66$, and outperforms the lower values. Moderate values like $\beta = 0.8$ (blue dashed line) and $\beta = 0.7$ (green dotted line) yield slightly lower rewards, stabilizing between $-67$ and $-68$. The lowest discount factor, $\beta = 0.6$ (purple dotted line), converges more quickly but levels off at a lower reward of approximately $-69$, indicating a bias toward short-term gains over long-term returns. This analysis suggests that higher discount factors (closer to 1) promote better long-term reward optimization, even if convergence is slower. WA3C performs best at $\beta = 0.9$, demonstrating its capacity to learn sustained value over time.

\medskip

\subsubsection{\textbf{Effect of Learning Rate \((\gamma)\) on \textit{WA3C} framework}}

\begin{figure}[htp]
    \centering
    \includegraphics[width=0.5 \textwidth, height = 2 in]{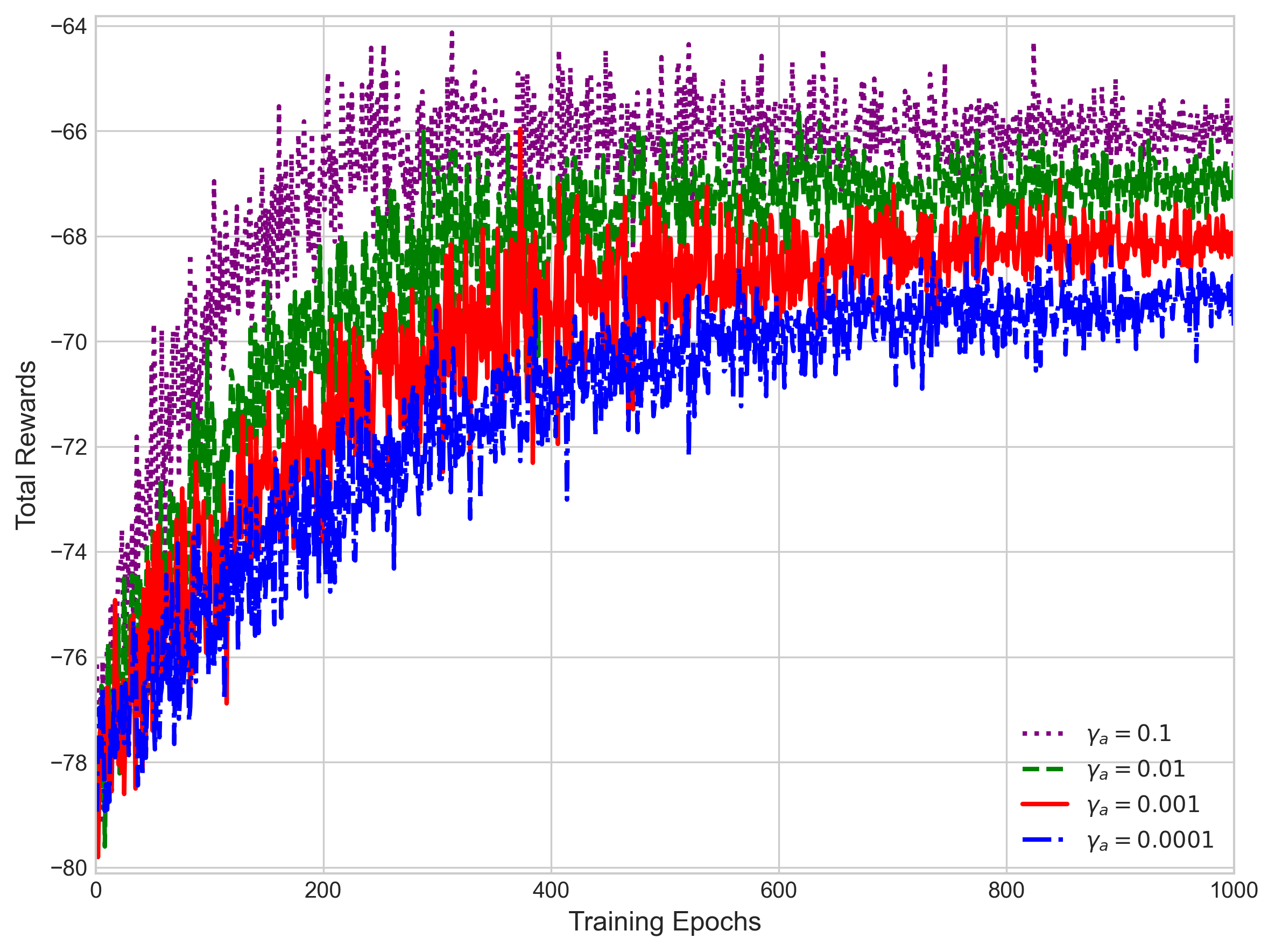}
    \caption{Sensitivity to varying Learning Rates}
    \label{fig:learning_rates}
\end{figure}

In \textbf{Figure~\ref{fig:learning_rates}}, we vary the learning rate across \(\gamma = \{0.1, 0.01, 0.001, 0.0001\}\)  to evaluate the trade-off between learning speed and stability. The moderate learning rate \(\gamma = 0.01\) (green dashed line) yields the best performance, achieving total rewards around \(-66.5\) with smooth and stable convergence. A higher learning rate, \(\gamma = 0.1\) (purple dotted line), shows rapid initial improvement but suffers from high variance and unstable convergence. In contrast, lower learning rates \(\gamma = 0.001\) (red solid line) and \(\gamma = 0.0001\) (blue dash-dotted line) converge slowly and stabilize at lower rewards of approximately $-70$ and $-72$, respectively. These results highlight that extreme learning rates can either destabilize the training or slow down learning significantly. A moderate learning rate \(\gamma = 0.01\)  strikes the optimal balance between convergence speed and training stability for \textit{WA3C}.

From Figs. \ref{fig:discount_factors} and \ref{fig:learning_rates}, we can say that \textit{WA3C} is most effective when configured with a higher discount factor (\(\beta = 0.9\)) and a moderate learning rate (\(\gamma = 0.01\)). These values ensure that the agent optimally balances short- and long-term gains while maintaining stable and efficient convergence.

\subsection{\textbf{Comparison with Non-RL Baselines}} 
Traditional non-learning-based scheduling algorithms such as Round Robin (RR), Shortest Job First (SJF), Longest Job First (LJF), Tetris, and Random scheduling are used as baseline references to evaluate the benefits of learning-based methods. These heuristics are widely adopted due to their simplicity and deterministic behaviour. By comparing the proposed \textit{WA3C} model against these baselines, we aim to assess the value added by incorporating adaptability and learning into the scheduling process, particularly in environments characterised by uncertainty and resource contention. 

\begin{figure*}
\begin{center}
$\begin{array}{cc}
\includegraphics[width=.4\linewidth,height=1.5in]{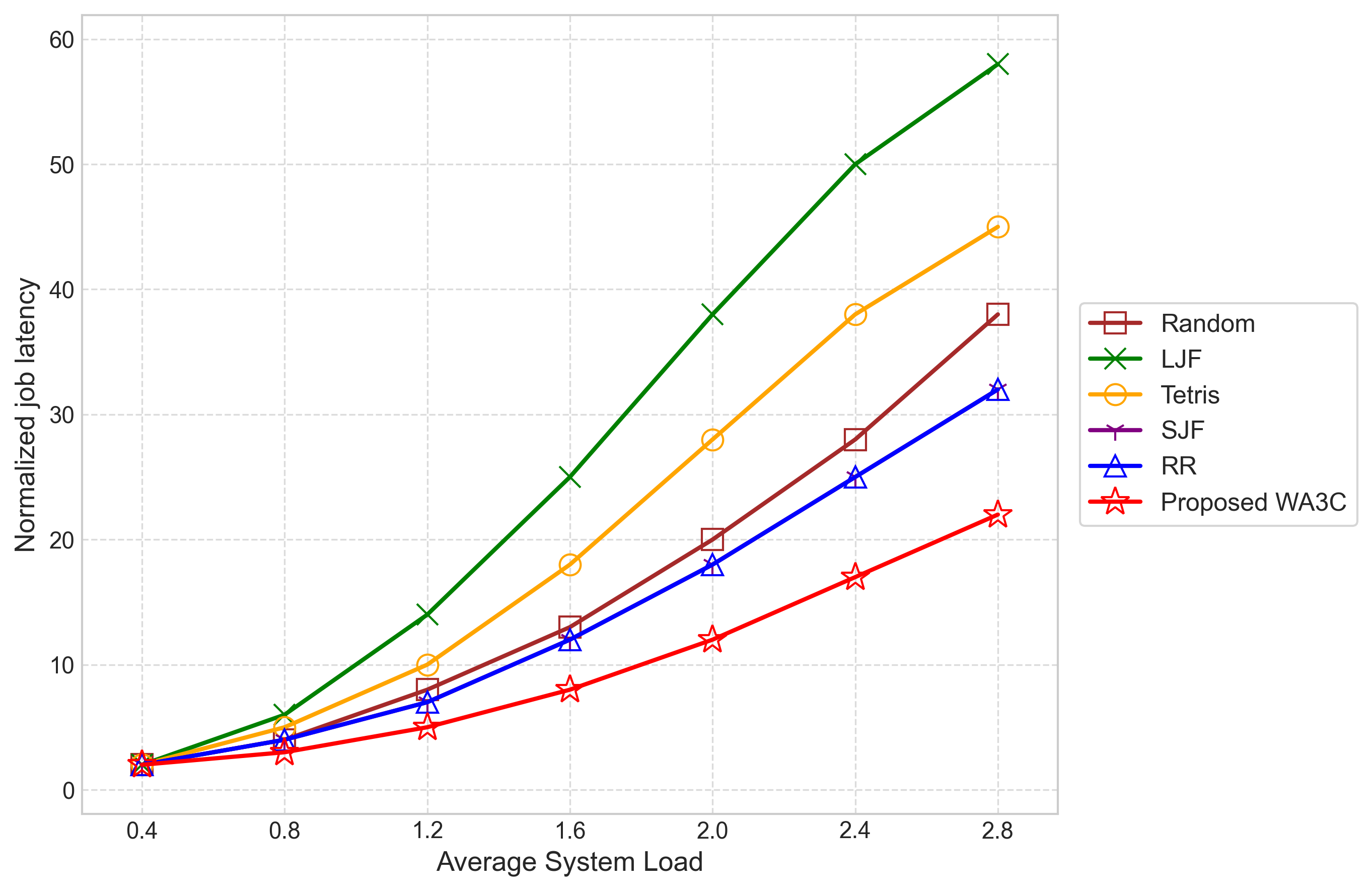}  &
\includegraphics[width=.4\linewidth,height=1.5in]{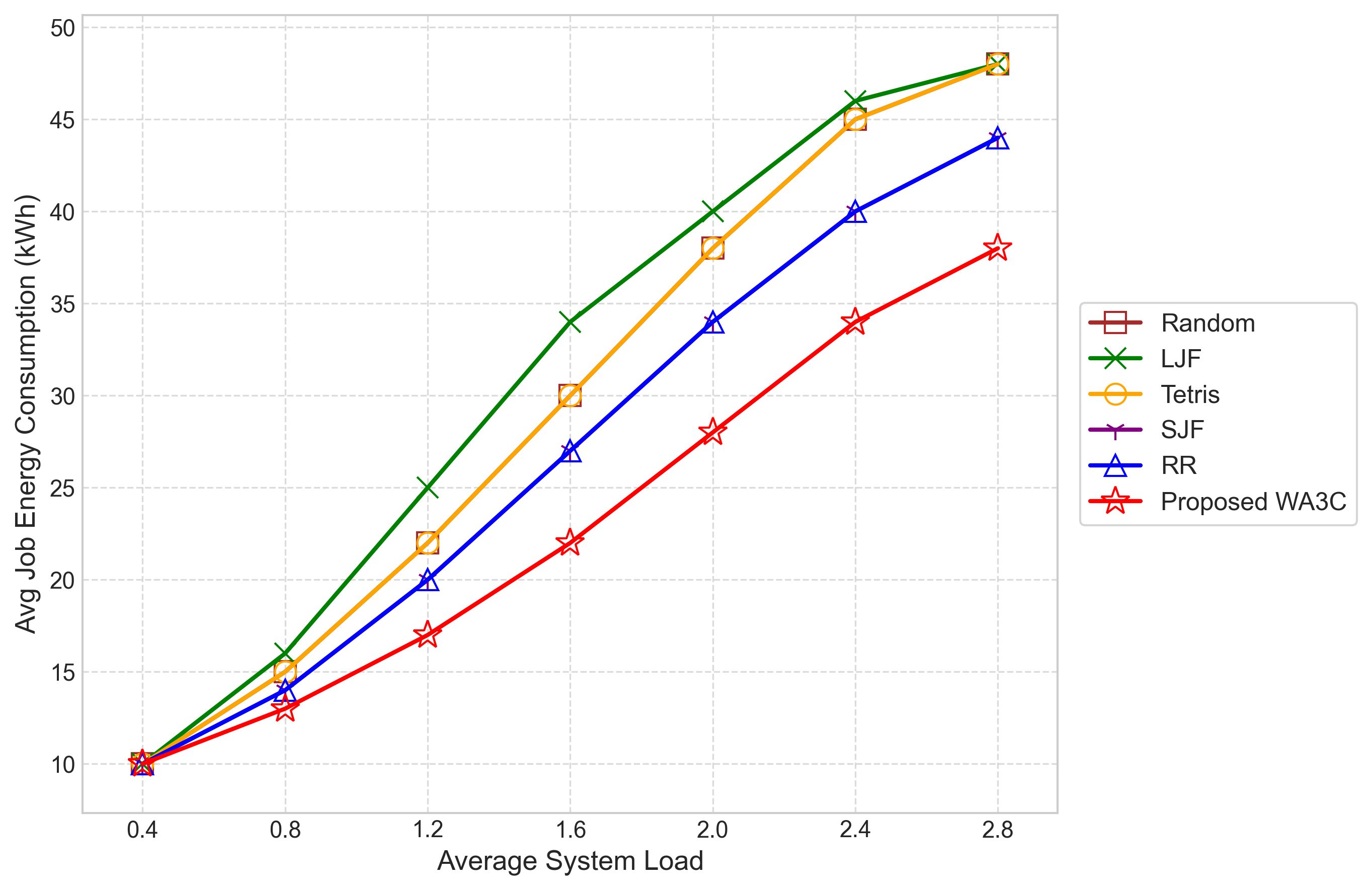} \\
\mbox{(a)} & \mbox{(b)} \\
\includegraphics[width=.4\linewidth,height=1.5in]
{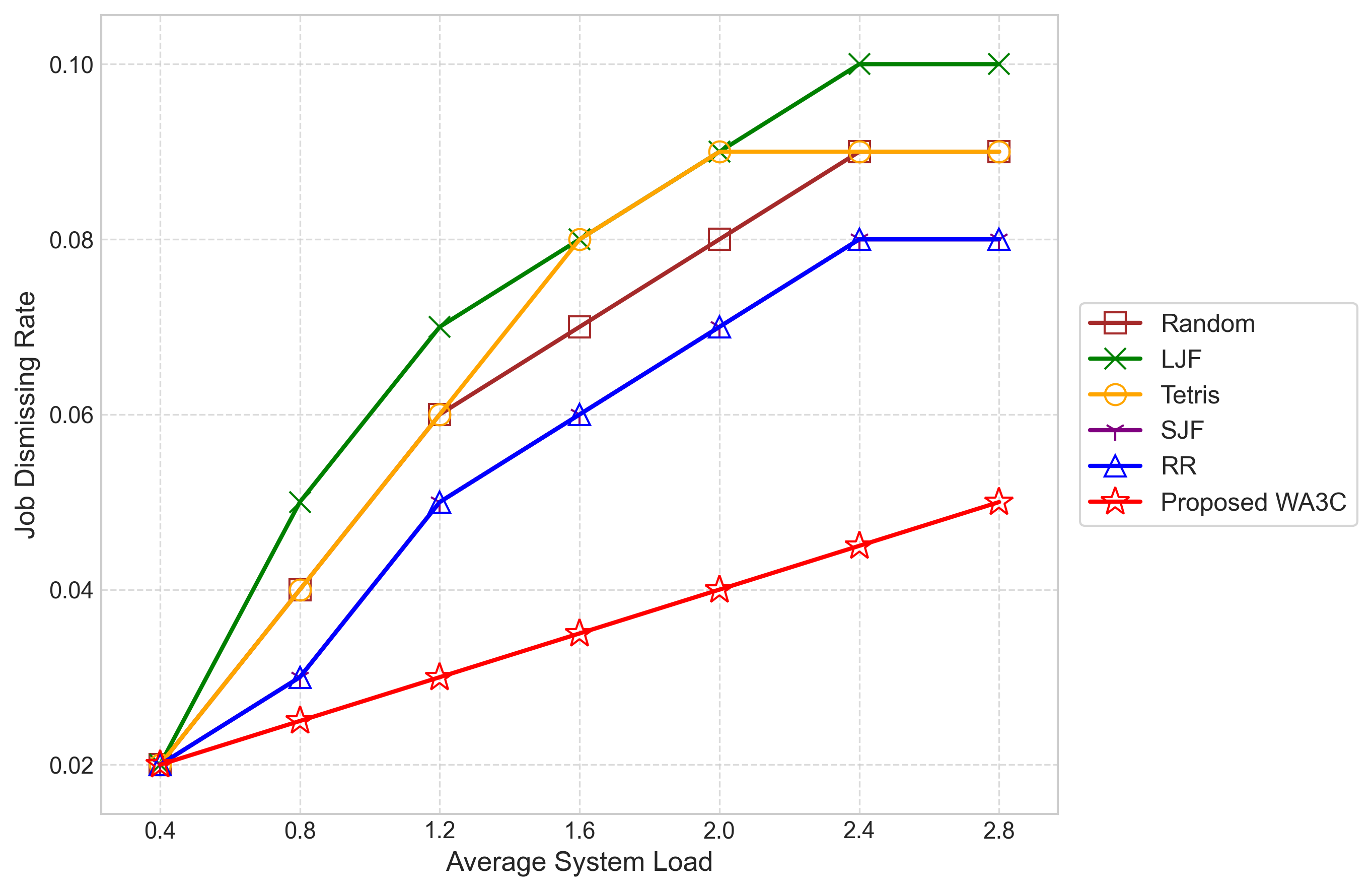}  &
\includegraphics[width=.4\linewidth,height=1.5in]{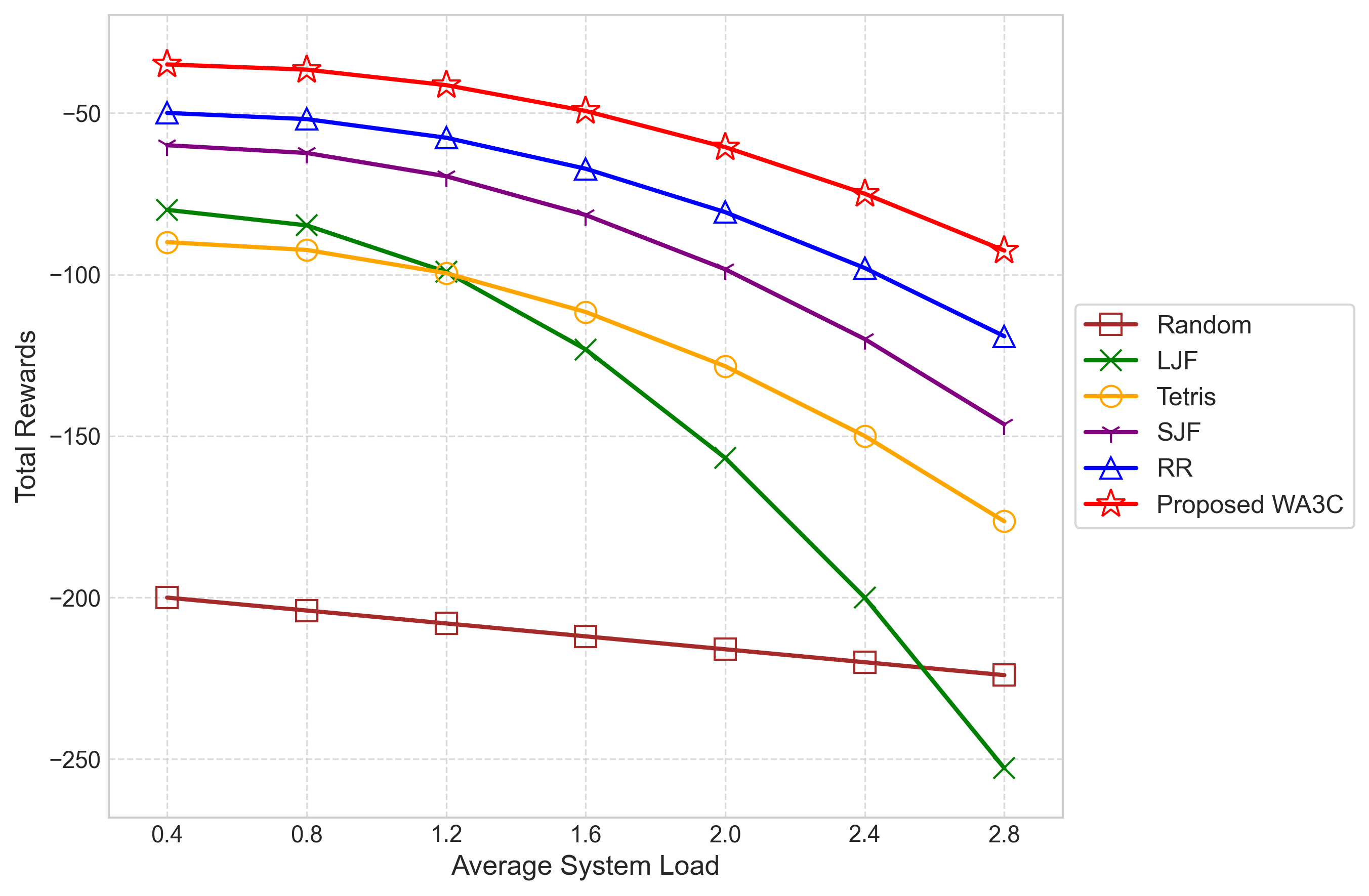} 
\\
\mbox{(c)} & \mbox{(d)} \\
\end{array}$
\caption{Comparison with non-Reinforcement Learning models considering (a) Average Job Latency, (b) Average Energy Consumption, (c) Job Dismissal Rate, and (d) Total Reward per Episode}
\label{Rewire_example}
\end{center}
\end{figure*}

The performance of each technique is analysed using the four key evaluation metrics across increasing average system load levels (from $0.4$ to $2.8$), as shown in \textbf{Figure [\ref{Rewire_example}]}. This allows us to systematically evaluate the robustness and scalability of WA3C as the scheduling environment becomes progressively more complex and congested.

\begin{enumerate}[{I.}]
\item{\textbf{\textit{Average Job Latency:}}} In \textbf{Figure~\ref{Rewire_example}(a)}, \textit{WA3C} consistently achieves the lowest latency, maintaining values between $2$ and $22$ units. In contrast, \textit{LJF} shows the highest latency, sharply increasing to $58$ due to its tendency to prioritize long jobs. \textit{Tetris} peaks at $45$, while \textit{SJF} and \textit{RR} stabilize around $32$. \textit{Random} also underperforms, with latency rising to $38$. \textit{WA3C}’s ability to adaptively prioritize tasks enables it to maintain low waiting times even under increasing load, demonstrating an effective and dynamic response to network congestion.

\item {\textbf{\textit{Average Energy Consumption:}}}
\textbf{Figure~\ref{Rewire_example}(b)} illustrates that \textit{WA3C} achieves the lowest job energy consumption, increasing gradually from $10$ to $38$ \textit{kWh} as load intensifies. In contrast, \textit{LJF} and \textit{Tetris} consume up to \textit{48 kWh} due to inefficient scheduling and overlapping jobs. \textit{SJF} and \textit{RR} perform moderately, stabilizing around \textit{44 kWh}, while \textit{Random} exhibits a similarly poor profile to \textit{LJF}. \textit{WA3C}’s superior energy efficiency results from its ability to intelligently batch and schedule jobs, minimizing idle power usage and avoiding hardware overutilization.

\item{\textbf{\textit{Job Dismissal Rate:}}} As shown in \textbf{Figure \ref{Rewire_example}(c)}, \textit{WA3C} maintains the lowest dismissal rate curve, peaking at $0.09$ even under maximum system load. \textit{LJF} reaches $0.10$, showing increased starvation of short jobs. \textit{Tetris} and \textit{Random} both plateau at $0.09$ while \textit{SJF} and \textit{RR} cap at $0.08$. 
\textit{WA3C}’s contextual awareness enables it to prioritise resource allocation to high-utility jobs, preventing job drops even under resource contention.
\item{\textbf{\textit{Total Reward per Episode:}}}
\textbf{Figure \ref{Rewire_example}(d)} demonstrates how each method accumulates rewards across training epochs. The proposed \textit{WA3C} begins with a total reward of $-35$ and exhibits the slowest rate of degradation (quadratic slope $10$), outperforming all baselines. In contrast, \textit{Random} scheduling starts at $-200$ and linearly drops with slope $10$, quickly becoming unviable under higher loads. \textit{LJF} and \textit{Tetris} decay steeply (slopes $30$ and $15$ respectively), while \textit{SJF} and \textit{RR} degrade moderately. \textit{WA3C}’s performance peak and consistent trend under growing load signify its superior long-term learning and adaptability.

\end{enumerate}

From Fig. \ref{Rewire_example}, it can be inferred that WA3C consistently outperforms all non-RL baselines across every metric, showcasing both \textit{effectiveness} (higher rewards and lower latency) and \textit{stability} (energy efficiency and low job dismissal). This validates the benefit of reinforcement learning in adaptive and high-load scheduling scenarios. Importantly, WA3C achieves these performance gains while dynamically respecting job \textit{priorities} and maintaining \textit{fairness} across diverse workloads, ensuring that critical tasks are favoured without starving lower-priority jobs.

\subsection{\textbf{Comparison with DRL Methods}}  
\begin{figure*}[t]
\begin{center}
$\begin{array}{cc}
\includegraphics[width=.4\linewidth,height=1.75in]{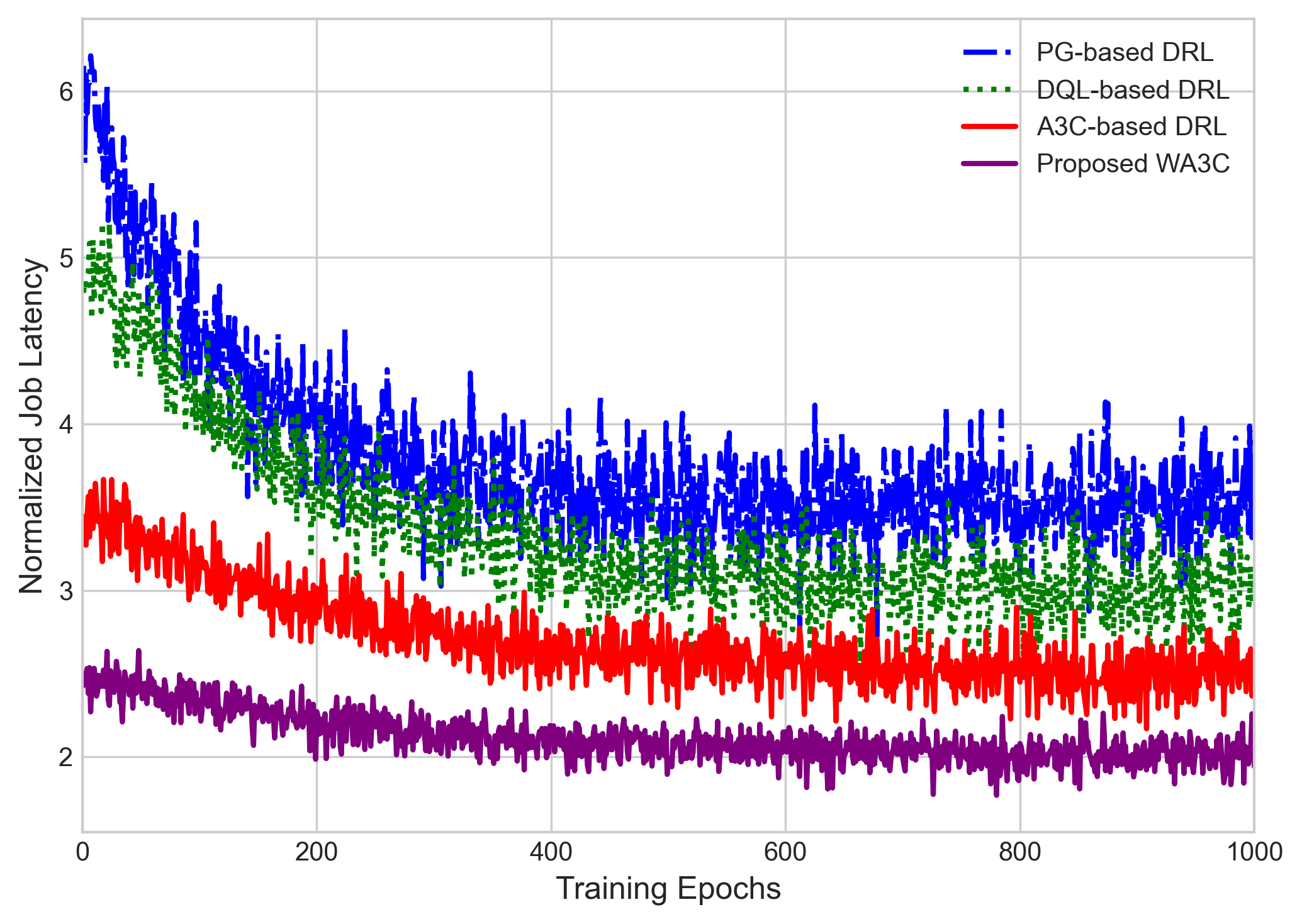}  &
\includegraphics[width=.4\linewidth,height=1.75in]{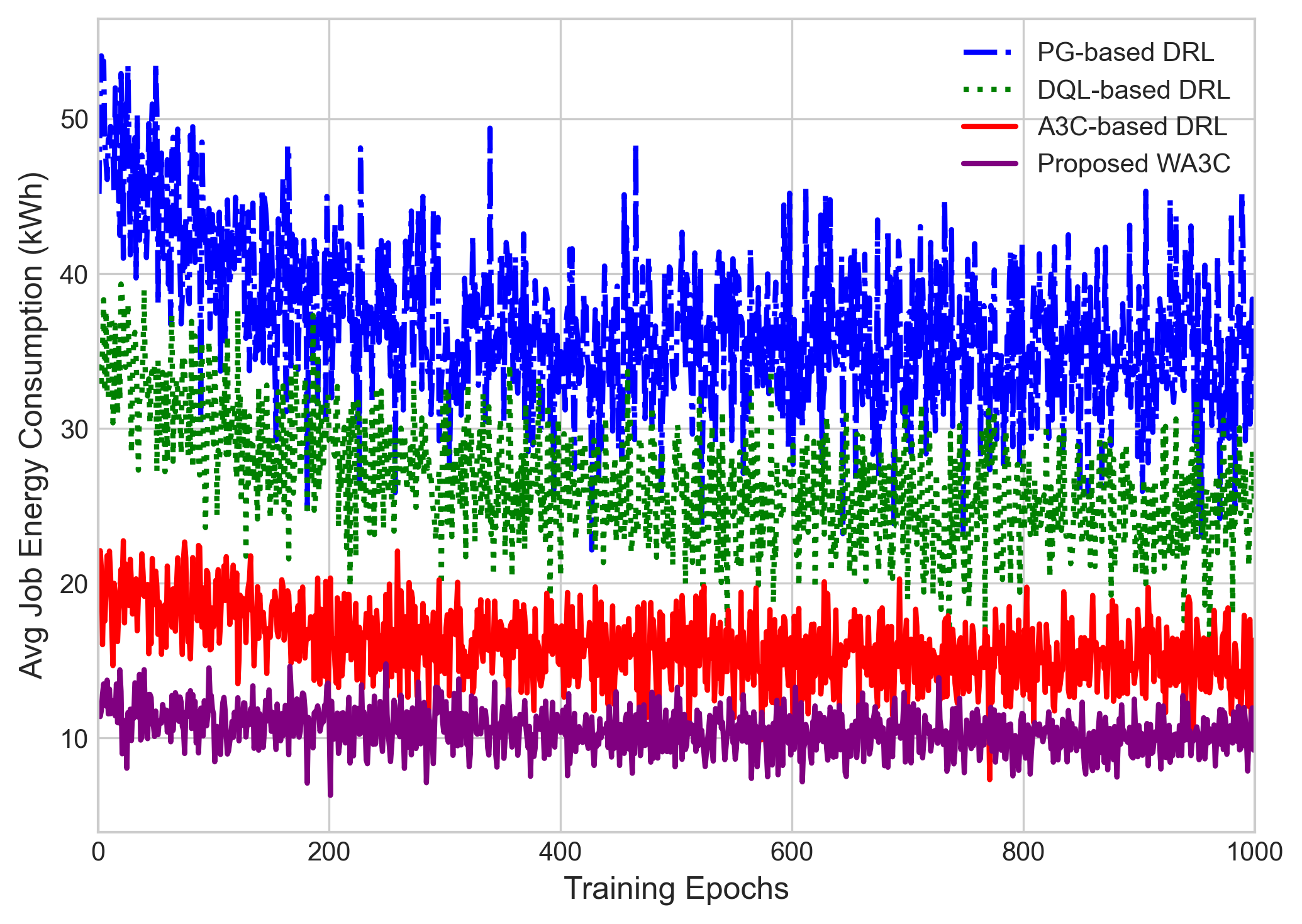} \\
\mbox{(a)} & \mbox{(b)} \\
\includegraphics[width=.4\linewidth,height=1.75in]{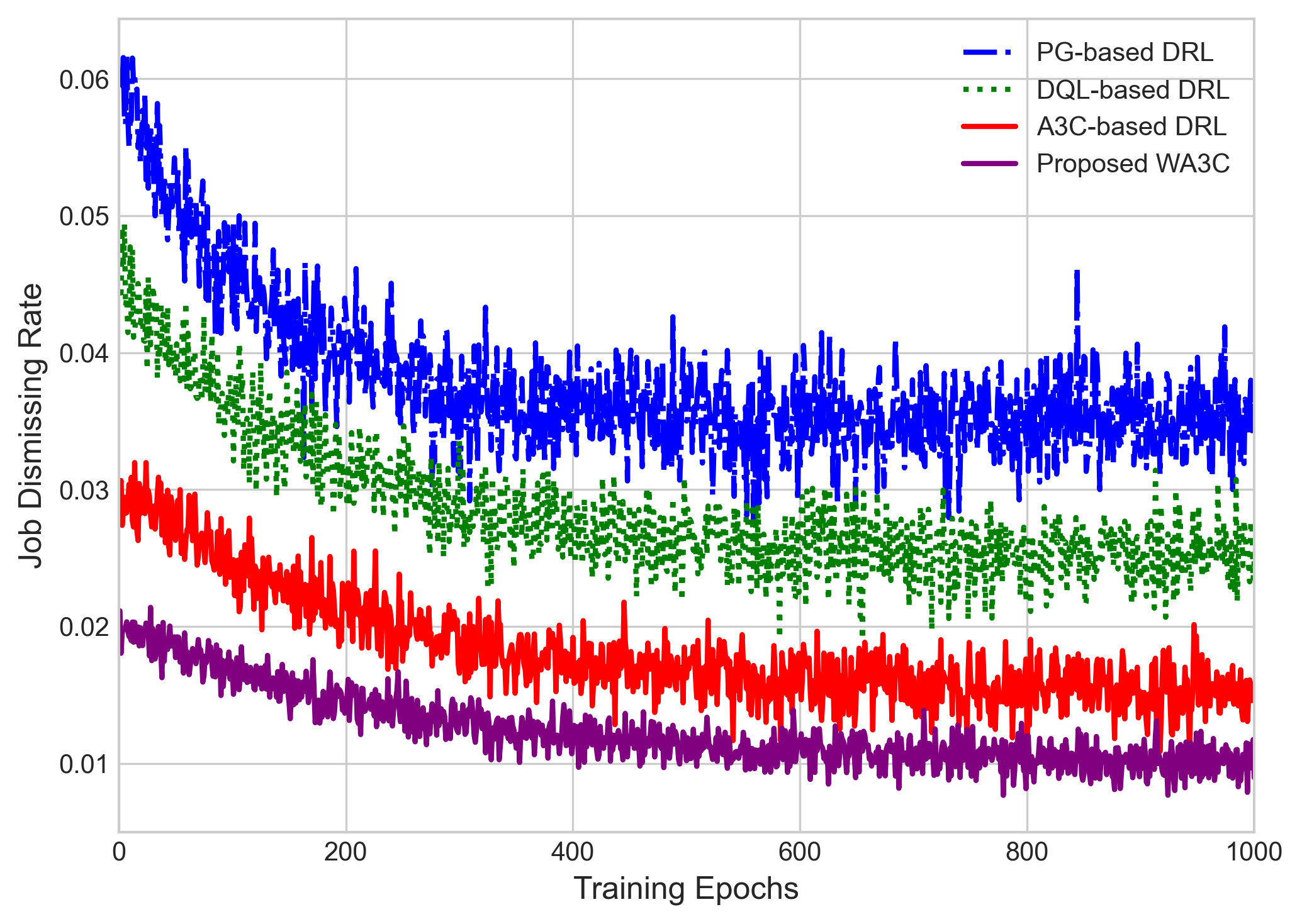}  &
\includegraphics[width=.4\linewidth,height=1.75in]{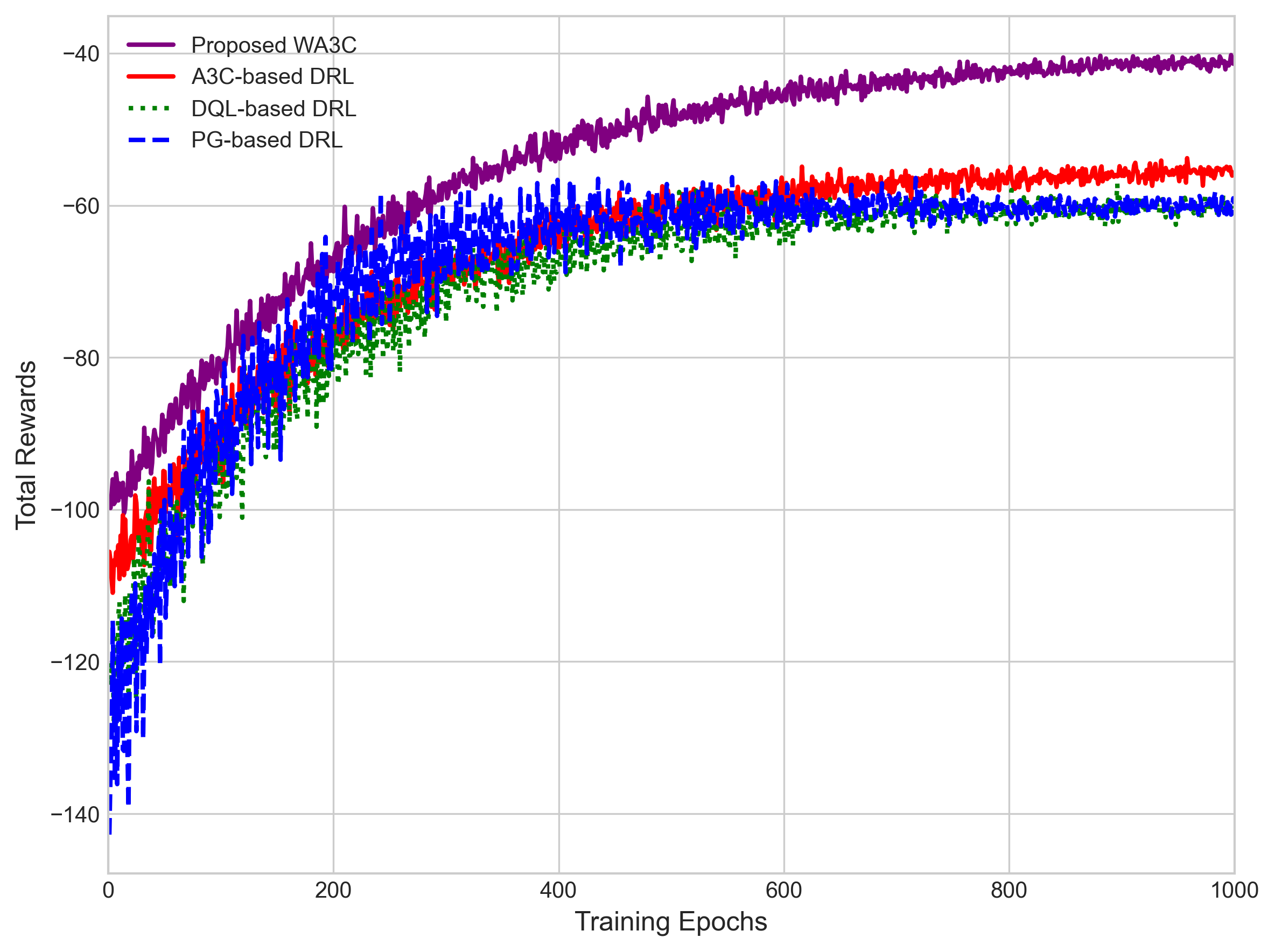} 
\\
\mbox{(c)} & \mbox{(d)} \\
\end{array}$
\caption{Comparison with Deep Reinforcement Learning models considering (a) Normalized Job Latency, (b) Average Energy Consumption, (c) Job Dismissal Rate, and (d) Total Rewards per Episode.}
\label{Rewire_example1}
\end{center}
\end{figure*}

To understand WA3C’s relative standing among reinforcement learning approaches, we compare it with other prominent DRL methods such as \textit{Asynchronous Advantage Actor Critic(A3C), Deep Q-Learning (DQL), and Policy Gradient (PG)}. These methods represent diverse learning paradigms (value-based, policy-based, actor-critic). This comparison enables us to isolate the contribution of WA3C’s architectural modifications and weighting strategy, and to evaluate how well it balances learning efficiency, convergence speed, and reward stability in complex scheduling tasks. These models are trained under identical simulation settings for fair comparison. Performance is analysed using four key metrics across training epochs, as shown in \textbf{Fig. \ref{Rewire_example1}}.

\begin{enumerate}[{I.}]
    \item{\textbf{\textit{Normalized Job Latency:}}} \textbf{Fig. \ref{Rewire_example1}(a)} shows the latency trends across epochs. \textit{WA3C} maintains the lowest latency curve, stabilizing near $2$. \textit{A3C-based DRL} achieves latency of $2.5$, while \textit{DQL} and \textit{PG} degrade to $3.0$ and $3.5$, respectively. \textit{WA3C}’s superior responsiveness stems from its adaptive prioritisation and efficient job placement.
    
    \item{\textbf{\textit{Average Energy Consumption:}}}
    \textbf{Figure \ref{Rewire_example1}(b)} demonstrates that \textit{WA3C} consistently outperforms others with energy consumption starting at $12~kWh$ and converging to $10~kWh$. \textit{A3C-based DRL} uses slightly more energy (down to $15~kWh$), while \textit{DQL} and \textit{PG} remain at $25~kWh$ and $35~kWh$ respectively. This validates \textit{WA3C}’s hardware-aware scheduling strategy, avoiding over-utilisation and idle power waste.
    \item{\textbf{\textit{Job Dismissal Rate:}}} \textbf{Figure \ref{Rewire_example1}(c)}: \textit{WA3C} achieves the lowest dismissal rate, converging to $0.01$. \textit{A3C} and \textbf{DQL} plateau at $0.015$ and $0.025$, respectively. \textit{PG} remains the highest, peaking at $0.035$. \textit{WA3C}’s ability to sustain fairness and prioritise critical jobs results in minimised job losses under load.
    \item{\textbf{\textit{Total Rewards:}}}
    In  \textbf{Fig. \ref{Rewire_example1}(d)}, the proposed \textit{WA3C} achieves the highest cumulative rewards throughout training. \textit{WA3C} begins at $-65$ and quickly improves, asymptotically approaching $-55$ with a slow decay governed by $\exp(-\frac{epochs}{250})$. \textit{A3C-based DRL} follows with a reward curve from $-68$ to $-56$, showing decent convergence. \textit{DQL-based DRL} starts lower at $-75$ and converges more slowly due to higher policy instability. \textit{PG-based DRL} performs the worst, with initial rewards at $-80$ and only minor improvements, saturating at $-60$. $WA3C$ demonstrates superior long-term learning and reward optimisation, highlighting its advantage in dynamic scheduling.
    \end{enumerate}

From the results in Fig. \ref{Rewire_example1}, we can say that the proposed \textit{WA3C} model consistently outperforms all \textit{RL}-based baselines across total reward, job latency, energy efficiency, and job dismissal rate. These improvements are a direct result of \textit{WA3C}’s novel integration of workload-aware state representation, adaptive action selection, and asynchronous actor-critic updates tailored for dynamic cloud environments. Unlike standard DRL models such as \textit{A3C, DQL}, and \textit{PG}, which primarily focus on maximising cumulative rewards without explicitly addressing job-level fairness or priority constraints, \textit{WA3C} incorporates both fairness and job criticality directly into its policy optimisation. This enables it to prioritise urgent and high-impact jobs while still maintaining overall system balance.

\section{\textbf{Discussion}} \label{Sec5}
This section presents a discussion of the proposed \textit{WA3C} model, its practical implications, and the limitations of the strategy.
\subsection{\textbf{Analysis of WA3C Model's Effectiveness}}

This work proposes \textbf{Weighted A3C (WA3C)}, a novel reinforcement learning-based framework that builds upon \textit{A3C} model to enable intelligent job scheduling in real-time cloud computing environments. The proposed approach addresses critical challenges such as fairness, energy efficiency, job urgency, and Quality-of-Service (QoS) compliance by incorporating a dynamically weighted multi-objective reward function. Traditional scheduling algorithms often fail to balance these competing metrics, whereas \textit{WA3C} provides a flexible and adaptive mechanism to tune system behaviour through learned policy updates.

A key innovation of this work is the introduction of a composite reward function, which linearly combines five major components: QoS adherence, energy consumption minimisation, priority handling, fairness (using Jain's Index), and penalties for job dismissals. This allows the agent to learn a scheduling policy that not only maximises throughput but also maintains system-level objectives aligned with operational goals. Additionally, the action selection strategy is improved by modifying the standard softmax policy to integrate job priority using a tunable parameter $\alpha$, enabling a trade-off between selecting high-priority jobs and maintaining fairness across the job pool.

\begin{figure}[htp]
    \centering
    \includegraphics[width=8cm]{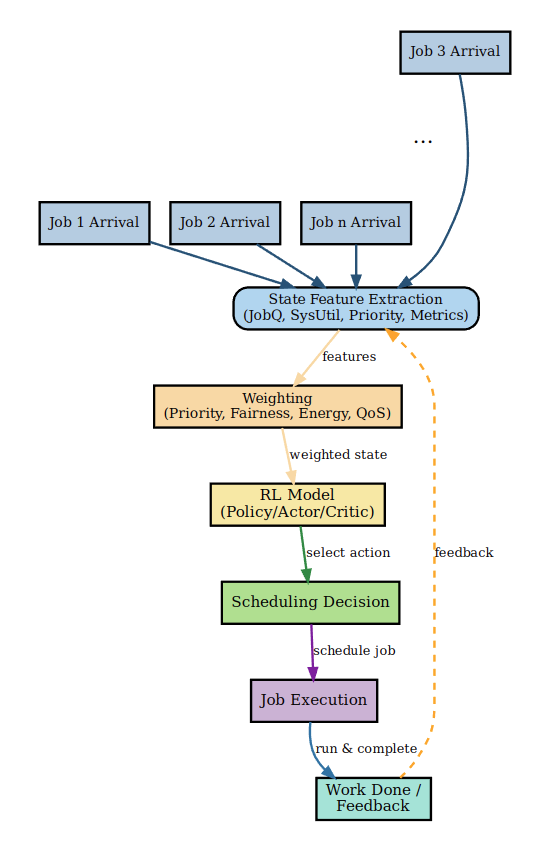}
    \caption{Flow of execution and collaboration between agents in the\textit{ WA3C }model. Each worker independently observes the environment, takes actions, and updates shared global parameters asynchronously.}
    \label{fig:Flow of Execution}
\end{figure}

The flow of execution in the \textit{WA3C} framework is illustrated in Fig. \ref{fig:Flow of Execution}. It begins with each asynchronous agent observing the current system state, which includes the job queue, system load, timing constraints, and prior outcomes. Based on this input, the actor network proposes an action namely, the assignment of a job to an appropriate resource using the modified softmax strategy that incorporates both estimated Q-values and job priorities. The selected action is then executed in the cloud environment, leading to updates in system state variables and the generation of a composite reward signal. This reward is derived from the weighted sum of multiple objectives as defined in the reward function. Subsequently, the agent computes the Temporal Difference (TD) error, reflecting the gap between predicted and actual value estimates. This error guides updates to both the actor (policy) and critic (value) networks through backpropagation. Each agent learns independently in parallel threads and periodically synchronises with a shared global network by pushing gradients and pulling updated parameters. This asynchronous update mechanism accelerates convergence and enhances robustness in the face of non-stationary and variable workloads.

Furthermore, latency is modelled as the cumulative sum of queueing delay, waiting time, and execution time. This latency measure acts as an implicit signal within the reward, incentivising the system to reduce end-to-end job processing time. By combining weighted multi-objective modelling, priority-aware action selection, and asynchronous parallel learning, the \textit{WA3C} framework effectively adapts to dynamic cloud workloads while ensuring fair, energy-efficient, and QoS-compliant resource allocation.

\subsection{{\textbf{Practical Implication}}}

The proposed \textit{WA3C} model offers several significant practical benefits for cloud computing environments. By integrating job priority and fairness constraints directly into the reward function, \textit{WA3C} enables more efficient and adaptive resource allocation, ensuring high-priority tasks are serviced promptly while avoiding starvation of lower-priority workloads. This leads to improved Quality of Service (QoS) across a diverse range of applications, from real-time processing to large-scale data analytics. Furthermore, the model's dynamic learning capabilities allow it to respond effectively to changing system loads and workload patterns, making it well-suited for deployment in scalable, multi-cloud infrastructures. By promoting fairness and preventing resource monopolisation, the \textit{WA3C} approach supports more sustainable and balanced cloud operations, aligning with the increasing need for equitable and energy-efficient computing solutions.
\subsection{{\textbf{Limitation and Challenges}}}
 While the proposed \textit{WA3C} framework demonstrates improved performance in terms of efficiency, fairness, and Quality of Service, it is not without limitations. First, the training of deep reinforcement learning models like \textit{WA3C} requires substantial computational resources and time, which may hinder real-time adaptability in highly dynamic environments. Second, the design of an effective reward function that balances competing objectives, such as priority, fairness, energy efficiency, and performance, remains a challenging task and may require fine-tuning across different application domains. Finally, integrating \textit{WA3C} into existing cloud orchestration frameworks may demand non-trivial modifications to system architecture, posing a challenge for practical deployment at scale.

\section{\textbf{Conclusion and Future Scope}} \label{Sec6}

In this work, we presented \textit{WA3C}, a novel Deep Reinforcement Learning framework that extends the standard A3C architecture to enable adaptive, energy-efficient, and fair resource allocation in cloud data centers. \textit{WA3C} introduces explicit modeling of job prioritization and fairness constraints within the reward function, and incorporates a dynamic energy consumption estimation mechanism to account for operational cost factors typically overlooked by baseline approaches.

Extensive evaluations on synthetic job traces with varying priority distributions demonstrate that \textit{WA3C} consistently outperforms both traditional non-RL scheduling heuristics and contemporary DRL techniques across multiple performance metrics. Compared to non-learning-based baselines such as Round Robin, Shortest Job First, and Tetris, \textit{WA3C} offers significantly higher adaptability and responsiveness in dynamic environments, reflected by its superior reward accumulation and fairness enforcement. Furthermore, relative to standard DRL methods like A3C, Deep Q-Learning, and Policy Gradient, WA3C achieves higher mean rewards with reduced variance and faster convergence, attributed to its enhanced gradient aggregation and environment modeling mechanisms. \textit{WA3C} exhibits smoother convergence behavior and greater training stability, particularly under high system load and contention scenarios. These findings position \textit{WA3C} as a scalable and principled foundation for multi-objective scheduling in dynamic cloud environments.

To address current limitations, future work will explore meta-learning and continual learning techniques to reduce training overhead and improve adaptability in dynamic cloud environments. We also plan to investigate automated and multi-objective reward shaping to simplify the design of balanced reward functions across diverse workloads. Additionally, efforts will focus on integrating WA3C into real-world, heterogeneous multi-cloud infrastructures using modular APIs and lightweight policy deployment mechanisms. Incorporating production-scale workload traces will further validate WA3C’s applicability and enhance its generalization in practical cloud orchestration scenarios.

\bibliographystyle{IEEEtran}
\bibliography{main}

\end{document}